# FINDING CLUSTERS OF SIMILAR-MINDED PEOPLE ON TWITTER REGARDING THE COVID-19 PANDEMIC

Philipp Kappus and Paul Groß

Department of Computer Engineering, Baden-Wuerttemberg Cooperative
State University, Friedrichshafen, Germany

## ABSTRACT

*Two clustering methods to determine users with similar opinions on the Covid-19 pandemic and the related public debate in Germany will be presented in this paper. We believe, they can help gaining an overview over similar-minded groups and could support the prevention of fake-news distribution. The first method uses a new approach to create a network based on retweet-relationships between users and the most retweeted users, the so-called influencers. The second method extracts hashtags from users posts to create a "user feature vector" which is then clustered, using a consensus matrix based on previous work, to identify groups using the same language. With both approaches it was possible to identify clusters that seem to fit groups of different public opinions in Germany. However, we also found that clusters from one approach cannot be associated with clusters from the other due to filtering steps in the two methods.*

## 1. INTRODUCTION

During the years of 2020 and 2021 the Covid-19 pandemic and the resulting legal regulations polarized the society and led to a huge discussion dominating the social networks, including Twitter (most used hashtag in Germany of 2020: "corona" [1]). In Germany, the so called "Querdenker" (eng.: lateral thinkers) gained nation-wide attention by organizing protests against Covid-19 related regulations, partly denying the existence of the virus and spreading various conspiracy theories on Twitter.

The aim of this work was to identify groups of similar opinions on this topic by analysing the tweets posted on Twitter in Germany regarding the Covid-19 pandemic. To achieve this, a novel method was developed, establishing a communication network based on relations between users, where one retweeted the other similar to [2]. A new way of filtering the data regarding only connection to influential users was used to distil clusters from the otherwise chaotic and over-connected network. This led to the possibility of grouping "normal users" to a so-called "superuser", if they retweet the same influencers. The DBSCAN algorithm is then used to detect communities inside the filtered network.

Furthermore, an implementation of the previously published hashtag-based clustering method of [3], where a combination of several rounds of the k-means algorithm and the DBSCAN algorithm is proposed, has been applied to the dataset. The latter approach applied to the tweets about the pandemic and its visualization yields insight into the different hashtags used by different groups in the debate.

Results from both methods can be viewed on https://andfaxle.github.io/twitteranalysis/.

## 2. LITERATURE REVIEW

Twitter, with its 192 million daily active (monetizable) users [4] expressing their views and opinions in short texts, has been a popular playground for data research analysing peoples interest in topics or products, performing personality studies and even predict flu outbreaks [5]. Previous work in this area can be divided into two groups of clustering approaches: Network Clustering and Content Clustering. While Network Clustering establishes a graph between users with retweets, mentions or followers connecting them, Content Clustering uses NLP to analyse the actual text that have been posted in terms of keywords used, sentiment or various other parameters.

[6] clustered the content of tweets by hashtags with the k-means algorithm, agglomerative hierarchical clustering and a fuzzy neighbourhood model. Similar to that, [3] analysed 30,000 tweets from just before a world cup to cluster the content in order to extract topics from the tweets. To reduce noise, they proposed four different algorithms one of which runs several rounds of the k-means algorithm with varying $k$ using the cosine distance on keywords extracted from the tweets. A so-called consensus matrix is then created stating in how many rounds two users end up in the same cluster. Users that have been clustered together in more than 50 per cent of the k-means rounds, can now be considered a community.

[7] used a combination of both content and network approaches. They developed a classifier grouping users into political-left and political-right by first establishing a network of reference users based on retweets (and mentions). Afterwards they assigned features for each group, consisting of keywords extracted from the users tweets in the cluster.

[2] examined the influence of Russian trolls in the context of #BlackLivesMatter by setting up a retweet network that clearly showed two distinct clusters (political-left, political-right) and continued to analyse the influence of trolls on each of these clusters.

## 3. DATA ACQUISITION

Since the terms of service are accepted, users agree that Twitter can make their content "available to other companies, organizations or individuals" [8] by providing access over the Twitter API. An AWS architecture was used to automatically retrieve and store tweets regarding the Covid-19 pandemic. An EC2 instance (t2) running a Python script registered on Twitters filtered-stream API to retrieve tweets in german language and containing the keywords "covid" and "corona". Through a Kinesis Data Firehose, the tweets are stored as bundles in a S3 - Bucket. Over the course of March 2021, a total of 2,955,282 tweets posted by 260,954 different users were collected.

## 4. NETWORK CLUSTERING

Users can be regarded as nodes of a graph. A relationship between two users can be interpreted as edges and is established when one of them retweets the other. The approach to create a complex graph, has been used several times as noted in section 2 by [7] or [2]. Both regard all retweets as valuable connection, but this makes the graph unfeasible large and complex. We propose to regard only connections to "influencers" in order to distil a subgraph that is nearly as meaningful but offers great and more nuanced insights into communities in the Twitter landscape. The whole process was implemented using Python orchestrating basic file and folder operations, no external database or tools have been used.

### 4.1. Influencers

Analysing the data set, we found that 31.54 per cent of all retweets are originally posted by the same 100 users (0,04 per cent of all users), who, as of now, will be called influencers. We can therefore assume, that these influencers primarily shape the opinion-landscape and are core users of possible clusters. A graph $G = (N, E)$ can be build regarding only relationships between users $U$ and the 100 influencers $I$ with edges $(u, i, g) \in E$ and $g$ representing how often the user $u$ retweeted the influencer $i$. Testing with a subset of four days and only three influencers using matplotlib and an implementation of the NEATO-layout algorithm [9] takes 35.2 seconds on a standard linux computer and produces a graph depicted in figure 1.

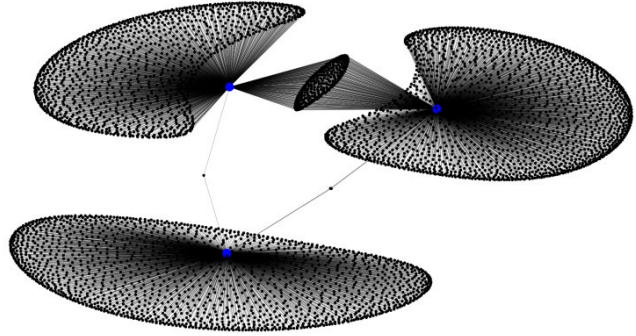

Figure 1: Graph using three influencers (blue)

### 4.2. Superusers

As seen in figure 1, there are many users only retweeting one of the influencers and some retweeting two. To further reduce complexity, we can aggregate all users that retweet the same influencers to one superuser $s$.

$$S_{i^n .. i^m} = \{u \in U \mid \forall (u, i^n .. i^m) \in E\] \quad (1)$$

This limits the number of possible nodes in the graph to:

$$|K| = |S| + |H| = 2^{|H|} - 1 \quad (2)$$

The weights $g$ of the edges are summed up.

### 4.3. Thresholding

With 100 influencers there are still more than $10^{30}$ possible nodes. To further minimize complexity a threshold is used to cut edges with weights lower than a threshold $T$:

$$E_{filtered} = \{(s, i, g) \in E \mid g > T\} \quad (3)$$

A good method of determining the best threshold value has not been established but using 0.65 per cent of the maximum weight in the graph produces clear clusters.

### 4.4. Normalizing weights

While analysing the count of retweets on the 100 most retweeted users (influencers), we found out that they are distributed according to an inverse power law:

$$N_{retweets} = b * x^{-m}$$
$$m = 0.65$$
$$b = 39,215$$

Where $x$ is the rank, $m$ is the slope and $b$ is the scaling factor or the number of retweets of the influencer at rank 0. Following the nature of an inverse power-law and as seen in figure 2, the higher ranks have significantly greater retweet counts, making it hard to find a threshold that on the one hand minimizes complexity regarding connections to the higher ranked influencer while keeping groups consisting of lower ranked influencers in the graph. To counteract this problem, weights of edges to influencers are multiplied by the common logarithm of the rank of this influencer before applying the threshold. This does not completely normalize the retweets as it would overvalue retweets of lower ranked influencer but decreases the dominance of higher ranked influencer.

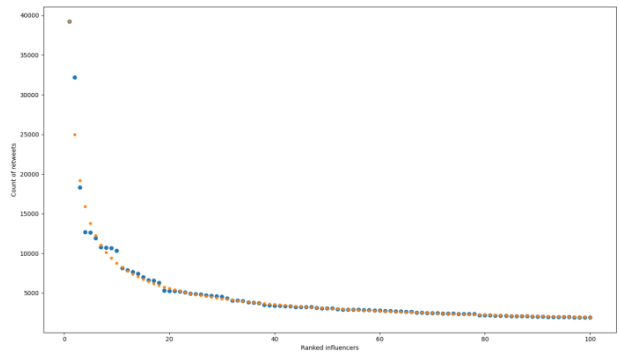

Figure 2: The distribution of retweets (blue) and the power-law (orange)

### 4.5. Clustering

These steps applied on the full dataset of one month, 100 influencers and a threshold of 61 took 2 hours and 58 minutes to calculate and yielded a graph depicted in figure 3.

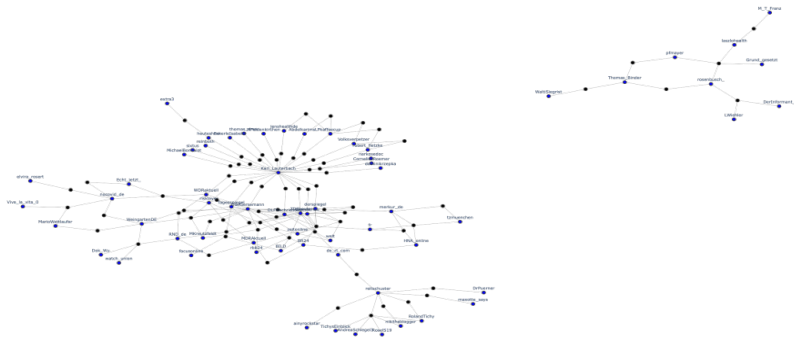

Figure 3: Full graph from one month and |I| = 100 and T = 61

A human can identify a main cluster in the middle, a smaller one connected to the main one on the bottom and a separated cluster on the upper right. To automatically identify these, a DBSCAN algorithm was used where a core influencer is defined if it is connected to $minPts = 2$ other influencers (over superusers). A minimal distance ε does not have to be defined since by applying a threshold, irrelevant connections are already filtered out. In order to prevent the algorithm to cluster all nodes that are in any way connected (the lower cluster has one connection to the main cluster) the algorithm is modified in such a way, that already visited nodes do not count as a new neighbour, reducing the number of core points. The resulting clustered graph can be seen in figure 4 and in the appendix.

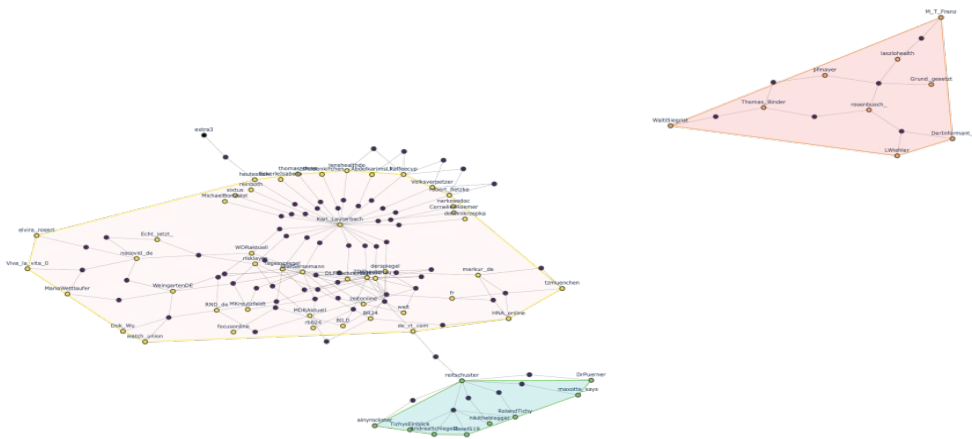

Figure 4: The same graph as in figure 3 but clustered using DBSCAN

### 4.6. Looking into the clusters

Let's have a look on some of the influencers of each cluster. The main cluster seen in yellow consists of public and private news agencies like "ZDF", "tagesschau", "derspiegel" or "BILD" as well as the most retweeted user in the dataset: "Karl Lauterbach". He is a present figure in the public debate on the pandemic based on his background in the social democratic party (SPD) and as a medical practitioner.

Some influencers from the green cluster are: "maxotte_says" (Max Otte) former leader of the "Werteunion" a faction within the Christian Democratic Union (CDU), known for advocating more conservative positions. "ainyrockstar" is a right-winged journalist [10] and "RolandTichy" former editor of "Impuls" and "Euro".

An influencer from the upper right, orange cluster is "rosenbusch" (Henning Rosenbusch), an independent journalist, advocating the "Swedish-way". Also, the user called "laszlohealth" (unknown) has a pinned tweet: "Corona has become a strange mixture of religious and political war by all means. The mask is the symbol of belonging. PCR mass testing the weapon. Objectivity, freedom of expression and normal interaction no longer exist." ( [11] translated with google translation). Another example is "Thomas Binder", a swiss doctor who has been advocating an anti-regulation position and was arrested and admitted to psychiatry because of suspected threats against politicians [12] and whose account has been blocked.

## 5. LANGUAGE CLUSTERING

In this approach we implement the algorithm of [3] discussed in section 2, assuming that users sharing the same opinion on the topic will also use the same hashtags. Summing these hashtags up, a map of hashtags can be assigned to each user. Creating a user feature vector and comparing them using the cosine distance, the k-means algorithm and the DBSCAN algorithm can be used to detect similar language preferences among users.

### 5.1. Hashtag Extraction and Preparation

Initially, hashtags are extracted from the individual tweet object as the Twitter API already delivers them in a separate field. For data preparation and to reduce variance among all extracted hashtags they are traced backed to their root word, aka lemmatized. This was implemented using the HanTa library for Python [13]. Furthermore, words that have no semantic value for the sentence and are only included for grammatical reasons (e.g.: "like", "the" etc.), are filtered out. These words are called "stop words" and are based on a detailed list of German stop words from [14].

### 5.2. User Feature Vector Creation

In order to find communities of users, a list associated to each user is created stating which hashtags he used and how often. Since the amount of all words across all users is very large, it must be reduced first. This involves a loss of information but is necessary to make the data processable. All hashtags which are present only once and whose overall count does not move within the 97 per cent and 99.98 per cent quantile, are removed. Thus, all hashtags which are not frequently used or are used by everyone (and therefore provide no information) are sorted out. The upper limit is necessary because the tweets were collected according to the hashtags concerning the Covid-19 pandemic (#covid, #corona, #covid-19) and are therefore contained in all tweets.

Making the list of hashtags comparable using the cosine distance, a binary feature vector is created for each user. The number of dimensions on this vector is equal to the count of distinct hashtags in the whole dataset. If a user has used a hashtags more than three times, a 1 is put in the dimension corresponding to this hashtag, otherwise a 0.

## 5.3. User Clustering

A well-established clustering method is the k-means algorithm. A variable $k$ is used to determine how many clusters a dataset should be divided into. The centers of each cluster are chosen randomly and every user is assigned to the cluster where the cosine distance to its center is lowest. In the next step new centers are calculated as the average characteristic values of all users contained in the cluster. Every user is then reassigned again. This is done until there are no more changes. As the number of clusters is predetermined by $k$ and the centers are initialized randomly, the algorithm is not optimal, since the number of clusters cannot be determined beforehand. Furthermore, all users are assigned to clusters, which makes it impossible to exclude noise.

As [3] proposed these issues can be solved in creating a consensus matrix by running k-means multiple times with different values for $k$. A null matrix of size $n \times n$ is created, where $n$ is the number of users and each row and column represent a particular user. For each run of k-means, the value within the matrix at (User A, User B) and (User B, User A) is increased by one if these users end up in the same cluster. The resulting matrix holds a value for each pair of users stating how often they ended up in the same cluster and therefore how similar the characteristics of the users are.

From the consensus matrix a graph can be imagined between all users where the value in the matrix defines the weight of the edge between these two users. To find communities of users that are densely connected (they used the same hashtags), the DBSCAN algorithm is used, where users are regarded as points. A core point is defined as a point that has $minPts$ of connected points with an edge-weight of ε or greater. Both are hyperparameters but setting $minPts$ to 2 per cent of the total users in the filtered dataset and ε to 80 per cent of the times k-means was run (the maximal possible value in the matrix) worked well. In the case of March 2021: $minPts = 20, \varepsilon = 15$. An advantage of the DBSCAN is that the aforementioned noise points are labelled as such. With the found clusters the graph was formed and visualized using Gephi and the ForceAtlas Layout Algorithm [15]. Figure 5 shows the graph using 15 iterations of k-means.

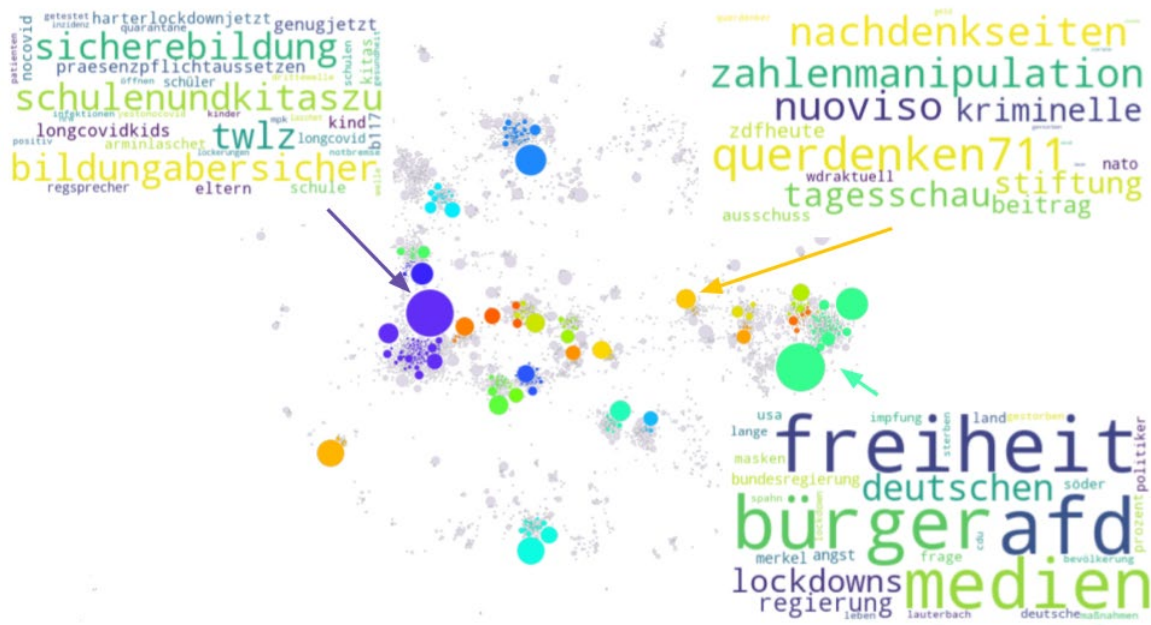

Figure 5: A graph created by evaluating the similarity of different user

## 5.4. Interpreting Clusters

In order to find the hashtags that constitute each cluster, we compared the relative frequency of a hashtag used within a cluster to the relative frequency of that hashtag in the whole data set. The resulting word clouds of a sample of three clusters is depicted in figure 6.

The first one features the right-wing AfD together with the hashtags "freiheit", "medien" and "bürger" (eng.: freedom, media and citizens).

In the second cluster, no unambiguous subject can be found.

An example hashtag from the third cluster is "nachdenkseiten" which refers to a german journalistic web page. In recent history it was labelled as a "Conspiracy ideological and / or right-wing open media" ( [16] translated with google translation). Further hashtags are "kriminelle" and "zahlenmanipulation" (eng.: criminals and manipulation of numbers).

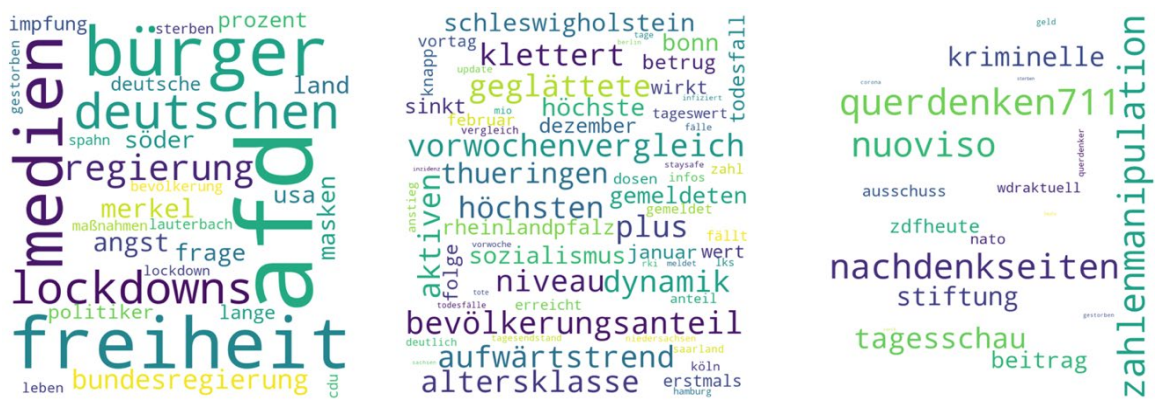

Figure 6: Word clouds of different user groups clustered by hashtag

## 6. COMPARISON OF CLUSTERING METHODS

To further consolidate the clusters found using both approaches, we compared the users belonging to each cluster in order to investigate whether it is possible to associate clusters from the retweet network approach to the language clustering approach. To illustrate the relationship a Sankey diagram (figure 7) was build depicting the network clusters on the left, language clusters on the right and users that are part of two clusters as a grey flow. Users from the network clusters, that do not belong to any language cluster flow to "undefined".

We found that more than 50 per cent of the users from the network clusters have not been clustered in the language approach (also vice versa). This can be explained by the steps in both approaches filtering users:

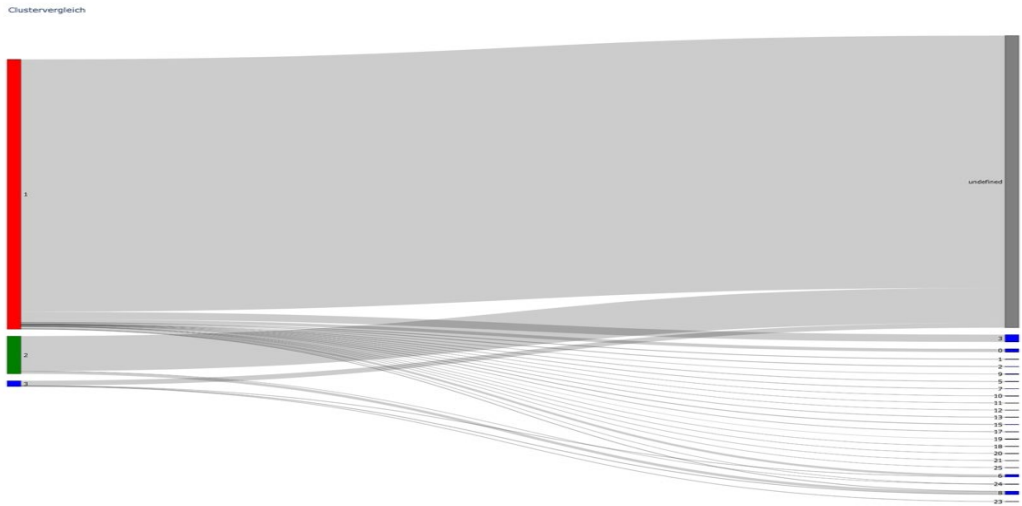

Figure 7: Comparison of users associations between clusters of the two approaches

**Filtering steps in the network approach**

The raw dataset holds 260,954 users.

- Only considering users that retweet: 37 per cent (966,322) users are filtered out.

- Only considering users that retweet influencers: 57 per cent of the users that retweet (94,730) are filtered out.

- Applying a threshold to superuser connections: Since some superusers will be deleted when all its connections are below the threshold their associated users are filtered from the cluster. 96 per cent of the users that retweet influencers (67,119) are filtered out.

**Filtering steps in the language approach**

- Only the hashtags whose counts are between the 97 per cent and 99.98 per cent quantile of the frequency distribution are used: 98 per cent (404,905) hashtags are filtered out.

- Only the users that have used those hashtags more than 6 times within the month are kept: 99.5 per cent (238,717) of all users are filtered out.

## 7. CONCLUSION AND FUTURE WORK

Twitter is a platform where millions of people publicly share their feelings and opinions every day. Analysing even large amounts of this information has become possible through the increase in computer power but also with the help of many dedicated research, coming up with better and better ways to structure this otherwise chaotic data. This paper presented two approaches to the problem of finding clusters in this unstructured data set. We showed that there are in fact clusters of users retweeting only themselves and using the same language regarding the debate on the Covid-19 pandemic. In developing a new approach and expanding methods already in place to distil this data and find clusters of similar-minded people we hope to distribute important information about the structure of the Twitter ecosystem and hope that further research can be conducted on top of our work.

Finding clusters in the retweet network heavily depends on the number of influencers and the threshold chosen. In future works, a method of choosing these parameters to reduce complexity only as much as necessary while keeping as many users in the data set as possible will mark a step ahead.

Expanding the language clustering method to keywords and fine-tuning the parameters for k-means and DBSCAN can yield clearer clusters. An important factor is the number of iterations of the k-means algorithm. More iterations with values for $k \geq 20$ or more would give more detail to the results.

Furthermore, since both methods used simple folder and file operations, a more sophisticated architecture could be set up to decreases computing time.

## BIBLIOGRAPHY

[1] H. Eberhardt, "Absatzwirtschaft," [Online]. Available: https://www.absatzwirtschaft.de/twitter-2020-ein-jahresrueckblick-mit-trends-und-hashtags-176871/. [Accessed 8 7 2021].

[2] L. Stewart, A. Arif and K. Starbird, "Examining trolls and polarization with a retweet network," 2018.

[3] D. Godfrey, C. Johns, C. Meyer, S. Race and C. Sadek, "A Case Study in Text Mining: Interpreting Twitter Data From World Cup Tweets," 2014.

[4] Twitter, Q1 2021 Letter to Shareholders, 2021.

[5] H. Achrekar, A. Gandhe, R. Lazarus, S.-H. Yu and B. Liu, "Predicting Flu Trends using Twitter data," 2011.

[6] G. Ifrim, B. Shi and I. Brigadir, "Event detection in Twitter using aggressive filtering and hierarchical tweet clustering," in CEUR Workshop Proceedings, 2014.

[7] M. Conover, B. Goncalves, J. Ratkiewicz, A. Flammini and F. Menczer, "Predicting the Political Alignment of Twitter Users," in IEEE Third International Conference on Privacy, Security, Risk and Trust and 2011 IEEE Third International Conference on Social Computing, 2011.

[8] Twitter, "Twitter Terms of Service," [Online]. Available: https://twitter.com/en/tos. [Accessed 8 7 2021].

[9] S. North, "Drawing Graphs with Neato," 2004.

[10] A. Graen, "Focus," [Online]. Available: https://www.focus.de/panorama/welt/panorama-anabel-schunke-ist-eine-der-wichtigsten-figuren-der-neurechten-szene-wir-waren-mit-ihr-feiern_id_10281656.html. [Accessed 8 7 2021].

[11] laszlohealth, "Twitter," [Online]. Available: https://twitter.com/laszlohealth/status/1319338449149874181. [Accessed 8 7 2021].

[12] Medinside, "Medinside," [Online]. Available: https://www.medinside.ch/de/post/verhafteter-aargauer-arzt-in-der-psychiatrie. [Accessed 8 7 2021].

[13] C. Wartena, " A Probabilistic Morphology Model for German Lemmatization," in 15th Conference on Natural Language Processing, 2019.

[14] J. Oppenlaender, "Github," [Online]. Available: https://github.com/solariz/german_stopwords/. [Accessed 8 7 2021].

[15]   M. Jacomy, T. Venturini, S. Heymann and M. Bastian, "ForceAtlas2, a Continuous Graph Layout Algorithm for Handy Network Visualization Designed for the Gephi Software," 2014.

[16]   M. Schwarzer, "Redaktionsnetzwerk Deutschland," [Online]. Available: https://www.rnd.de/panorama/esoteriker-auf-corona-demos-tanzende-hippies-neben-rechtsextremen-und-verschworungstheoretikern-was-will-sie-esoterikszene-QMRYIWRQLNCP5N5GGWMK7V53LM.html. [Accessed 8 7 2021].

**AUTHORS**

Philipp Kappus and Paul Groß are computer science students at the Baden- Wuerttemberg Cooperative State University in Friedrichshafen, Germany, spending half their study time at the campus and the other half at Airbus Defence and Space GmbH in Immenstaad. Due to the corona virus outbreak, both could only visit the courses online for more than one and a half years. That's why they became interested in analysing twitter to get insight on how different people think about and handle the pandemic.

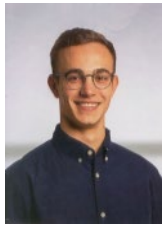
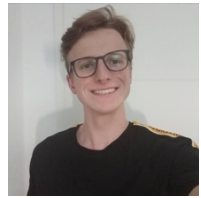